\documentclass[prl,twocolumn,showpacs]{revtex4}
\usepackage{amssymb}
\usepackage{epstopdf}
\usepackage{graphicx}
\usepackage{dcolumn}
\usepackage{amsmath}
\usepackage{amsfonts}

\begin{document}

%\preprint{}

\title[Short Title]{Forcibly driven coherent soft phonons in GeTe with 
intense THz-rate pump fields} 

\author{Muneaki Hase}
\author{Masahiro Kitajima}
\affiliation{Materials Engineering Laboratory, National Institute for 
Materials Science,1-2-1 Sengen, Tsukuba 305-0047, Japan 
}
\author{Shin-ichi Nakashima}
\affiliation{Power Electronic Research Center, National Institute of Advanced Industrial 
Science and Technology, 1-1-1 Higashi, Tsukuba 305-8561, Japan 
}
\author{Kohji Mizoguchi} 
\affiliation{Department of Applied Physics, Osaka City University, 
3-3-138 Sugimoto, Sumiyoshi-ku, Osaka 558-8585, Japan 
}
%\date{\today }

\begin{abstract} 
We propose an experimental technique to generate large
amplitude coherent phonons with irradiation of THz-rate pump
pulses and to study the dynamics of phase transition in GeTe
ferroelectrics. When a single pump pulse irradiates the sample at
various pump power densities, the frequency of the soft phonon
decreases sub-linearly and saturates at higher pump powers. By
contrast, when THz-rate pump pulse sequence irradiates the sample
at matched time intervals to forcibly drive the oscillation, a
large red-shift of the phonon frequency is observed without
saturation effects. After excitation with a four pump pulse
sequence, the coherent soft phonon becomes strongly damped leading
to a near critical damping condition. This condition indicates
that the lattice is driven to a precursor state of the phase
transition.
\end{abstract} 


\maketitle   

Femtosecond-attosecond laser technologies have recently been the
focus of much attention in solid state physics because of their
growing applications in observing both electronic and phononic
ultrafast dynamics. In particular, one of the possibilities for
the application of femtosecond laser pulses is controlling the
amplitudes of coherent collective motions of atoms excited in
condensed matter by using double pump pulses or multiple pump
pulses.\cite{Weiner90, Dekorsy93,Hase96,Ozgur01} One can enhance
the amplitude of a phonon mode by applying $in-phase$ pulses, or
suppress the amplitude by applying $out-of-phase$ pulses, both
phenomena are observed in a real time domain. The most remarkable
advantage of the multiple pulse pump technique is the ability to
avoid saturation effects due to high density excitation , i.e.,
screening of the space-charge field by electron-hole plasma.\cite{Liu96}

One goal in controlling coherent lattice vibrations is to cause
lattice instabilities which could lead to a phase transition. We
note that manipulation of phonons cannot be demonstrated with
conventional frequency-domain spectroscopy. Until now, all
experiments in coherent control of phonons have been performed
under low density excitation, in which the coherent phonon
amplitude was too small to observe lattice
instability.\cite{Weiner90, Dekorsy93,Hase96,Ozgur01} In this
letter, we propose an experimental technique to forcibly drive
{\it multiple} coherent phonons into {\it one} larger amplitude
coherent phonon and demonstrate its capability of inducing an
extremely unstable crystal phase close to the critical point in
ferroelectric materials. We used a twin Michelson interferometer
to produce intense femtosecond THz-rate pulse trains,
which were used to repetitively push lattice motions in-phase.
Thus, a larger amplitude coherent phonon was generated without
any saturation effects. The coherent phonon with larger
amplitude showed a large decrease in frequency and became 
strongly damped, showing that the lattice is driven into a 
strongly anharmonic regime near the structural phase transition.

We chose ferroelectric GeTe as a sample because of its
considerable interest in optical memories applications due to its
reversible structural change\cite{Okuda92}: This crystal undergoes
the rhombohedral-rocksalt structural change at the transition
temperature T$_{c}$ = 657$\pm$100 K, attributed to a displacive
phase transition.\cite{Steigmeiter70} In a displacive phase
transition, the motions between the two phases generally involve a
{\em soft mode} vibration, whose frequency is dramatically reduced
near T$_{c}$.\cite{Cochran59} This type of the phase transition is
characterized by a single potential minimum, whose position shifts
at T$_{c}$. By contrast, other type of the phase transition, such
as a order-disorder transition, is characterized by several
potential minima among which a choice is made at
T$_{c}$.\cite{Cochran59} The order-disorder transition occurs with
collective tunneling or thermally assisted hopping modes. In the
case of GeTe, the A$_{1}$ mode has been considered to be the soft
mode, which was observed by Raman
measurements.\cite{Steigmeiter70} An $ab$ $initio$ theoretical
investigation predicted that the phase transition in GeTe was a
fluctuation-driven first-order phase transition.\cite{Rabe87}

Because of the strongly reduced frequency of the soft mode near
T$_{c}$, monitoring the phase transition dynamics is difficult
with conventional frequency-domain spectroscopies.\cite{Cummins83}
Motivated by these difficulties, Nelson and coworkers examined
time-resolved pump-probe measurements at various lattice
temperatures in perovskites, and the heavily damped soft phonon
was observed near T$_{c}$.\cite{Dougherty92} The present study
approaches the critical point by increasing phonon amplitude
instead of increasing lattice temperature. The sample studied in
this work was a single crystal of GeTe prepared by a vapor growth
method and cleaved in the $c$ crystallographic plane. GeTe is a
narrow band-gap semiconductor, and the generation of the coherent
$A_{1}$ phonon is closely related to excitation of carriers from
the valence band to higher energy bands, i. e., displacive
excitation of coherent phonons (DECPs).\cite{Zeiger92,Forst00}
Femtosecond time-domain measurements were carried out at room
temperature with a pump-probe technique. Femtosecond laser pulses
of a Ti:sapphire laser, operating at 800 nm, were amplified to a
pulse energy of 500 $\mu$J in a 1 kHz regenerative amplifier.
After compensation of the amplifier dispersion, the amplified
pulses had 120 fs duration. The pump and probe beams were focused
on samples to a diameter of about 100 $\mu$m. The pump power
density was reduced by a neutral density filter to below 13
mJ/cm$^{2}$ to prevent damaging the sample or causing laser
ablation, and it was varied from 0.8 to 12.7 mJ/cm$^{2}$. The
probe pulse energy was also reduced and fixed at 0.3 mJ/cm$^{2}$.
The pump-beam was mechanically chopped at 315 Hz for the signal
detection by a lock-in amplifier. The transient reflectivity
change $\Delta R/R$ was recorded by changing the optical path
length of the probe beam.

\begin{figure}[ptb]
\includegraphics[width=7.5cm]{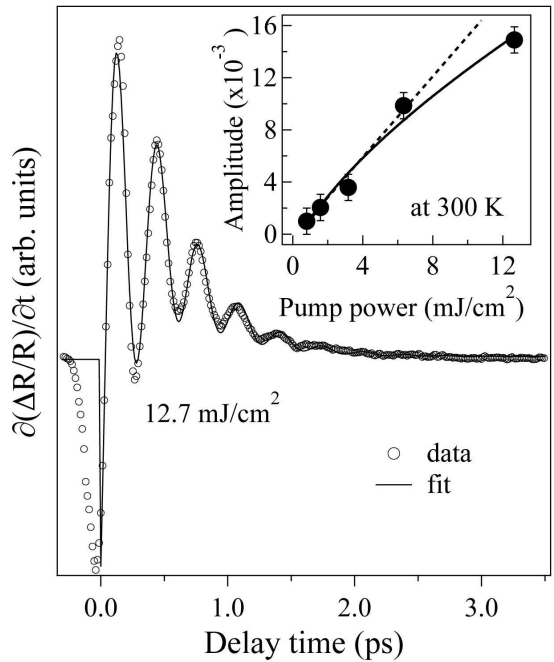}
\caption{The transient reflectivity change obtained at 12.7 mJ/cm$^{2}$. 
The open circles are the experimental data, the solid curve is the fit to the data 
with Eq. (1). 
The inset shows the amplitude of the coherent oscillation obtained by the fitting. 
} 
\label{Fig1}
\end{figure}
Figure 1 shows time derivatives of the transient reflectivity
change obtained by a single pump pulse at the pump power density
of 12.7 mJ/cm$^{2}$. The coherent oscillations due to the
collective motion of the crystal lattice appear on the slowly
varying background due to the photo-excited carriers. The
frequency and the amplitude were obtained by fitting the time
domain data using a damped harmonic oscillator with the
background,
\begin{eqnarray}
F(t) = Ae^{-t/\tau}cos(\omega t + \phi) + B[e^{-t/\tau_{1}} - e^{-t/\tau_{2}}],
\end{eqnarray}
where $A$ and $B$ are the amplitude of the coherent phonon and the
carrier contributions, respectively. $\tau$ is the dephasing
time of the coherent phonon, $\tau_{1}$ and $\tau_{2}$ are the relaxation
and rising times of the carrier background, respectively.
$\omega$ is the frequency and $\phi$ is the initial phase of the coherent
oscillation. The time period of the oscillation at the lowest pump power density
of 0.8 mJ/cm$^{2}$ is $\sim$ 263 fs
(= 3.80 THz), which is close to that of the $A_{1}$ mode observed by Raman
scattering, $\sim$ 3.81 THz (=127 cm$^{-1}$),\cite{Steigmeiter70} and
that for amorphous GeTe observed by coherent phonon spectroscopy.\cite{Forst00}
As the pump power density increases from 0.8 to 12.7 mJ/cm$^{2}$,
the time period of the $A_{1}$ mode increases, corresponding to the
red-shift of the phonon frequency from 3.8 to 3.0 THz, as discussed later.

As the power density of the single pump pulse increases, the
amplitude of the $A_{1}$ mode increases and saturates for the
highest employed fluence, as shown in the inset of Fig. 1.
Experiments were not performed above the fluence of 13 mJ/cm$^{2}$
because of sample damage by laser ablation through generation of
dense electron-hole plasma.\cite{Tinten98} A similar saturation
for the phonon properties was observed for the optical phonons in
semimetals\cite{DeCamp01} and in semiconductors\cite{Hunsche95}
under similar conditions of high-density single pump excitation. 
The phonon softening observed with the single pump pulse may be 
ascribed to the phonon self-energy effect\cite{Ledgerwood96} or 
the electronic softening.\cite{Hunsche95} The observation of 
underdamped oscillation with single pump pulse suggests that the 
GeTe crystal stays far from the critical point of the phase transition. 
In order to drive the $A_{1}$ mode closer to the critical point 
while avoiding saturation and sample damage, the multiple pump 
pulse excitation technique\cite{Wiederrecht95,Weiner00} was
applied to the $A_{1}$ mode.

\begin{figure}[ptb]
\includegraphics[width=8.8cm]{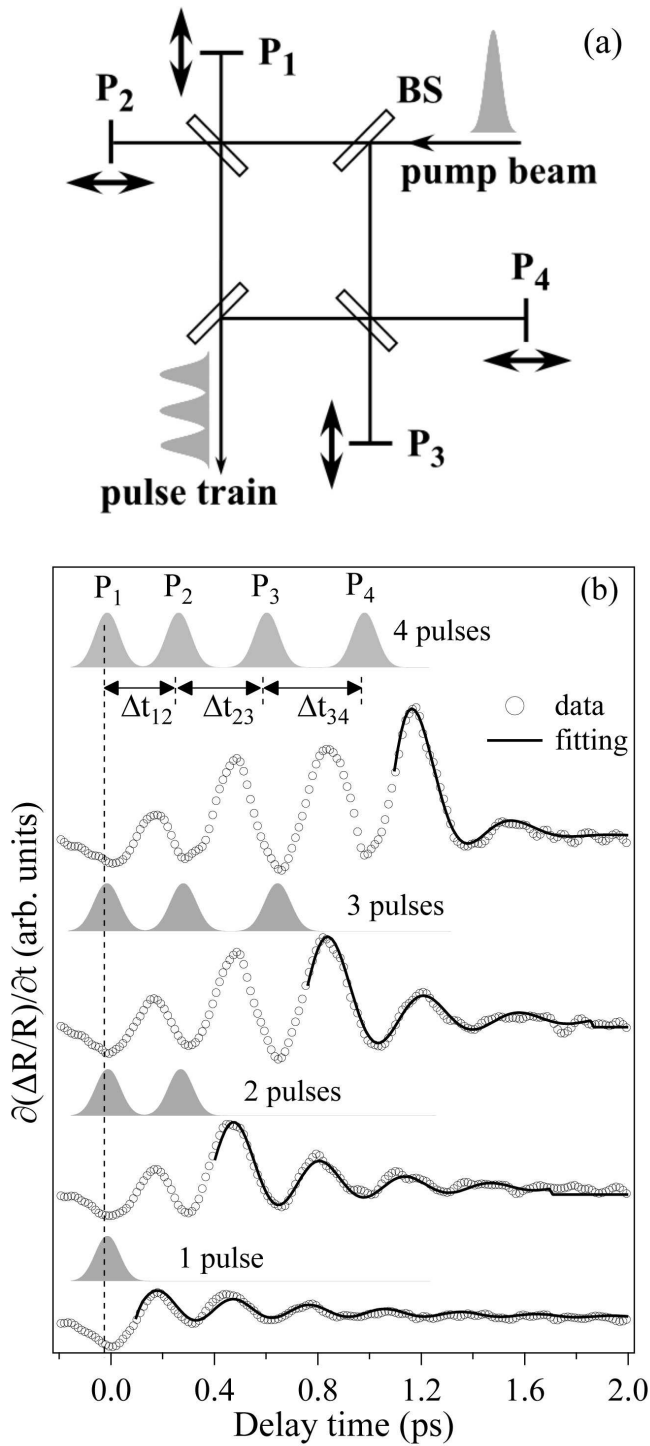}
\caption{(a) The optical layout of a twin Michelson interferometer for
the generation of the pulse train. BSs are the beam splitters. Each
mirror arm labeled P$_{i}$ (i=1, 2, 3, 4) is computer-controlled.
(b) Repetitive excitation of the $A_{1g}$ mode using THz-rate pulse
train. P$_{1}$, P$_{2}$, P$_{3}$, P$_{4}$ are first, second, third,
fourth pump pulses, respectively. $\Delta t_{12}$, $\Delta t_{23}$,
and $\Delta t_{34}$ are set to be 290 fs, 320 fs, and 345fs,
respectively. The open circles represent the experimental data.
The solid lines represent the fitting of the time-domain data after
the final pump pulse using Eq. (1).
} 
\label{Fig2}
\end{figure}
By adding two mirror arms to the Michelson
interferometer,\cite{Hase96} THz-rate pulse train with four pulse
sequence was generated at a variable repetition rate, as shown in
Fig. 2 (a). By moving the mirrors of arms in the twin
interferometer, the separation time $\Delta t_{ij}$ (i,j = 1, 2,
3, 4) between the pulse components of P$_{i}$ and P$_{j}$ was
controlled. The time derivatives of the transient reflectivity
changes obtained by using the multiple pump pulses are shown in
Fig. 2(b). Here, the power density of each pump pulse is 3.8
mJ/cm$^{2}$, such that the maximum total pump power (3.8
mJ/cm$^{2}$ x 4) exceeds that of the single pump excitation (12.7
mJ/cm$^{2}$) without sample damage. We aimed that each pump pulse
force the oscillation of the coherent $A_{1}$ mode through the
repetitive excitation. In order to well drive the soft phonon to
an in-phase motion, the time intervals in the THz-rate pulse train
was set to be unequal; as shown at the top of Fig. 2(b), the pulse
delay was increased for each subsequent pulse to match the
increased pulse intervals for each driven phonon oscillation. The
time-domain data in Fig. 2(b) clearly demonstrate both an
enhancement in the amplitude of the coherent $A_{1}$ mode by a
repetitive excitation and a drastic decrease in the dephasing time
of the coherent phonon. The data of Fig. 2(b) was fitted with Eq.
(1), in order to obtain the dephasing time of the coherent phonon
and the frequency of the $A_{1}$ mode for various total pump
powers.

\begin{figure}[ptb]
\includegraphics[width=7.0cm]{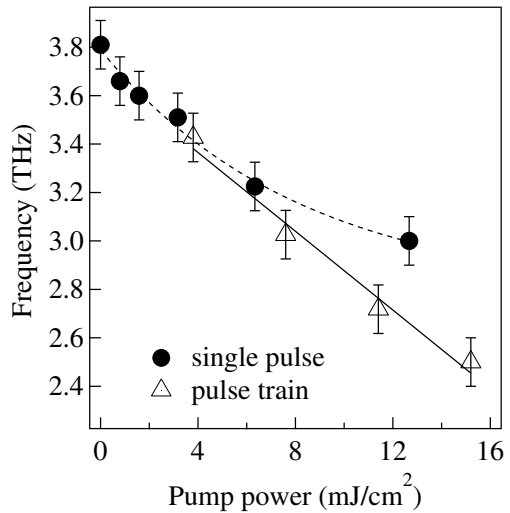}
\caption{The pump power dependence of the frequency of the $A_{1}$
mode obtained for the single pump pulse and the pulse train, respectively.
The dotted curve is an eye guide and the solid line corresponds to a fitting
of the data with a linear function. 
} 
\label{Fig3}
\end{figure}
Figure 3 shows the obtained frequency of the $A_{1}$ mode as a
function of the total pump power together with the results for
single pump excitation. The $A_{1}$ frequency decreases linearly
from 3.8 to 2.5 THz as we increase the number of pulse sequence,
while the single pulse excitation produce saturation at highest
fuences. The dephasing time decreases from 570 to 180 fs as the
number of pulse sequence is increased. The largest damping rate of
$\gamma$ $\sim$ 1.8 ps$^{-1}$ (=1/(180 fs)/$\pi$) observed with
four pump pulses is significantly close to the frequency of
$\omega$ $\sim$ 2.5 ps$^{-1}$ (or 2.5 THz), showing that the
$A_{1}$ mode is driven to near the point of the critical damping
($\omega$ = $\gamma$).

We note that the lowest frequency of the soft mode observed in the
present study using THz-rate pulse train, of 2.5 THz, is
comparable to the incoherent phonon frequency at a temperature of
$\sim$ 590 K, according to the Curie-Weiss law.
\cite{Steigmeiter70} This confirms that the crystal lattice is
really close to the transition point. 

In conclusion, we generated a large amplitude coherent phonon in
GeTe ferroelectrics by use of an intense THz-rate pulse train,
whose time period was matched to the increased phonon period. The
soft phonon frequency decreases linearly with the number of the
pulses in the excitation sequence, leading to near critical
damping condition for a 4 pulse excitation. The optical control of
the lattice vibration will initiate a new crystal structure, which
cannot be observed by conventional frequency methods. Such
experimental scheme for manipulating collective atomic motions can
be generally pursued in physics, biology, and chemistry.

The authors acknowledge T. Dumitrica and P. R. Vinod for critical reading 
of the manuscript. We thank the Venture Business Laboratory of Osaka 
University where the first stage of the pump-probe experiments were carried out.
The authors are grateful for the support received through a Grant-in-Aid
for the Scientific Research from the Ministry of Education, Culture,
Sports, Science, and Technology of Japan.

\end{document}